%% file: LaTeX/SUFFICIENT.tex
\documentclass[lettersize,journal]{IEEEtran}
\pdfoutput=1
\usepackage{cite}
\usepackage{amsmath,amssymb,amsfonts}
\usepackage{algorithmic}
\usepackage{graphicx}
\usepackage{textcomp}
\usepackage{enumitem} 
\usepackage{multirow}

\usepackage[dvipsnames,svgnames,x11names]{xcolor}

\begin{document}
\title{SUFFICIENT: A scan-specific unsupervised deep learning framework for high-resolution 3D isotropic fetal brain MRI reconstruction}
\input{LaTeX/section/author}

\maketitle

\input{LaTeX/section/abstract}

\input{LaTeX/section/introduction}
\input{LaTeX/section/method}
\input{LaTeX/section/exp}

\input{LaTeX/section/acknowledge}

\end{document}

%% file: LaTeX/section/author.tex
\author{Jiangjie Wu,  Lixuan Chen, Zhenghao Li, Xin Li, Saban Ozturk, Lihui Wang, Rongpin Wang, Hongjiang Wei, Yuyao Zhang 
\thanks{J. Wu, L. Chen, X. Li, Y. Zhang are with the School of Information Science and Technology, ShanghaiTech University, Shanghai, China (e-mail:
wujj@shanghaitech.edu.cn; chenlx1@shanghaitech.edu.cn;lixin22023@shanghaitech.edu.cn; zhangyy8@shanghaitech.edu.cn). }
\thanks{Z. Li, H. Wei are with the School of Biomedical Engineering, Shanghai Jiao Tong University, Shanghai, China (e-mail: lzh$\_$znso4@sjtu.edu.cn; hongjiang.wei@sjtu.edu.cn).}
\thanks{S. Ozturk is with the Department of Electrical-Electronics Engineering, Bilkent University, Ankara, Türkiye (e-mail:saban.ozturk@bilkent.edu.tr).}
\thanks{L. Wang is with the School of Computer Science and Technology, Guizhou University, Guiyang, China (e-mail: lhwang2@gzu.edu.cn).}
\thanks{R. Wang is with the Guizhou Provincial People's Hospital, Guiyang, China (e-mail: wangrongpin@126.com).}
}

%% file: LaTeX/section/abstract.tex
\begin{abstract}
High-quality 3D fetal brain MRI reconstruction from motion-corrupted 2D slices is crucial for clinical diagnosis.
Reliable slice-to-volume registration (SVR)-based motion correction and super-resolution reconstruction (SRR) methods are essential. 
Deep learning (DL) has demonstrated potential in enhancing SVR and SRR when compared to conventional methods. However, it requires large-scale external  training datasets, which are difficult to obtain for clinical fetal MRI. 
To address this issue, we propose an unsupervised iterative SVR-SRR framework for isotropic HR volume reconstruction. 
Specifically, SVR is formulated as a function mapping a 2D slice and a 3D target volume to a rigid transformation matrix, which aligns the slice to the underlying location in the target volume. 
The function is parameterized by a convolutional neural network, which is trained by minimizing the difference between the volume slicing at the predicted position and the input slice. 
In SRR, a decoding network embedded within a deep image prior framework is incorporated with a comprehensive image degradation model to produce the high-resolution (HR) volume. 
The deep image prior framework offers a local consistency prior to guide the reconstruction of HR volumes.
By performing a forward degradation model, the HR volume is optimized by minimizing loss between predicted slices and the observed slices.
Comprehensive experiments conducted on large-magnitude motion-corrupted simulation data and clinical data demonstrate the superior performance of the proposed framework over state-of-the-art fetal brain reconstruction frameworks.

\end{abstract}


%% file: LaTeX/section/introduction.tex
\section{Introduction}
\label{sec:introduction}

Fetal brain MRI is crucial for clinical prenatal diagnosis, offering a 3D structural volume and exceptional soft tissue contrast \cite{coakley2004fetal,cardenas2020fetal,davidson2021fetal}. Recognized for its reliability in quantitatively assessing cortical development and diagnosing congenital neurological disorders \cite{coakley2004fetal}, it serves as a valuable supplemental diagnostic tool in cases where ultrasound results are unclear, enhancing the accuracy of congenital disorder diagnoses and providing vital information about fetal development \cite{cardenas2020fetal}.
Acquiring a 3D high-resolution (HR) fetal MRI is challenging due to fetal and maternal body movement. Alternatively, fast 2D scanning processes with a large slice thickness (3-6 mm), are employed for efficiency. Multiple stacks at different orientations are acquired and combined to reconstruct an isotropic 3D fetal brain volume \cite{Girard2015,davidson2021fetal}. However, resulting 3D reconstructions suffer from inter-slice misalignment and low through-plane image resolution. A robust reconstruction method is crucial for enhanced brain pathology analysis and quantitative studies of brain development.

Existing reconstruction methods typically involve a two-step process: slice-to-volume registration (SVR) and 3D image volume super-resolution reconstruction (SRR)  \cite{alansary2017pvr,ebner2020automated,gholipour2010robust,hou20183,jiang2007mri,kainz2015fast,kim2009intersection,kuklisova2012reconstruction,rousseau2006registration,tourbier2015efficient}. First, multiple stacks of 2D fetal MRI slices are acquired in orthogonal orientations.
In each iteration, SVR registers these slices to a target volume for motion correction, followed by SRR to construct an HR 3D volume. The output volume becomes the SVR registration target for the next iteration. In these iterative approaches, the performance of both SVR and SRR significantly influences reconstructed image quality. However, conventional methods rely on good initial spatial consistency between slices acquired at different orientations, often failing to align motion-corrupted thick slices with large displacements. Subsequently, slice misalignment degrades the performance of the subsequent SRR.
Conventional SRR, for instance, tends to overfit to artifacts caused by motion corruption in MR images. If the initial SRR volume is significantly corrupted, the entire reconstruction process may fail to provide a globally optimal solution. Various efforts have been made to address these challenges. 

Supervised deep learning (DL) methods for SVR problem \cite{hou2017predicting,salehi2018real,ferrante2017slice,hou20183,singh2020deep} use convolutional neural networks (CNN) to learn a ground truth (GT) registration to map 2D slices to their corresponding locations in 3D space. While effective for highly misaligned slices, their accuracy may be insufficient for fetal brain MRI motion correction due to the lack of an appropriate reference volume.  Inference based on 2D slices resembles pose approximation via image retrieval rather than structure-based methods \cite{de2019deep,sattler2019understanding}. Efforts to improve SVR accuracy by incorporating features from 2D slices and 3D reference volumes \cite{leroy2022end,shi2022affirm,xu2022svort} encounter challenges. These include the requirement for an extensive high-quality training database, resulting in generalization issues. Additionally, they often overlook the blurring effect of the point spread function (PSF) during the image slicing procedure \cite{ebner2017point}, impacting the performance of the trained SVR model in clinical applications. An unsupervised SVR approach considering a physical forward model is highly desirable for improving motion correction.


Fortunately, VoxelMorph \cite{balakrishnan2019voxelmorph}, an unsupervised registration network, has shown remarkable efficiency in 3D volume-to-volume registration. 
Inspired by VoxelMorph, we propose a novel training dataset-free unsupervised SVR method for the random rigid motion correction of fetal brain MR images. 
We formulate the SVR task as a function that learns a parameterized rigid registration from an input 2D thick slice to its correct alignment in a 3D target volume. The SVR function is implemented using a CNN, which embeds image features of both slice and volume, then predicts a rigid transformation matrix aligning the images. The network is optimized by minimizing the difference between the volume slicing at the predicted position and the input 2D slice. 

Supervised DL methods for brain MRI SRR \cite{chen2018efficient,cherukuri2019deep,mcdonagh2017context,wang2020enhanced} excel in solving inverse problems, typically trained on LR-HR image pairs. However, applying these methods to HR reconstruction of fetal brain MRI poses challenges. The limited availability of a large high-quality training database hinders effective network training. Generalization issues emerge if input image parameters differ from those used in training.  Additionally, SRR methods are susceptible to image artifacts, and overfitting of these artifacts can introduce errors in the early iterative SVR-SRR step.

Recently, untrained deep generative models, like deep image priors (DIP) and related works \cite{cherukuri2019deep,sui2022scan,ulyanov2018deep,zhao2019channel}, leverage network architecture for robust image reconstruction. Despite the effectiveness of deep CNN-based generators, they often require early stopping to prevent overfitting artifacts. Addressing this, a recent approach introduced an under-parameterized deep decoder network for state-of-the-art (SOTA) performance \cite{heckel_deep_2018}. Inspired by this, we propose an unsupervised technique for 3D isotropic fetal brain SRR. Utilizing a decoding network with a comprehensive physical forward model, we map fixed noise onto a 3D volume. The under-parameterized deep decoder, establishes a robust MRI prior, reducing network overfitting in the presence of image artifacts.

We introduce SUFFICIENT, \itshape{a Scan-specific unsUpervised deep learning Framework For hIgh-resolution 3D isotropiC fetal braIn MRI rEcoNstrucTion}\upshape, to improve the robustness of HR fetal brain reconstruction. SUFFICIENT incorporates the aforementioned unsupervised SVR and SRR approaches to reliably perform 3D volume reconstruction from motion-corrupted stacks. 
Experiments conducted on large-magnitude motion-corrupted simulation data and clinical data demonstrate that SUFFICIENT has superior performance compared to the SOTA fetal brain reconstruction frameworks. 
Our contributions are primarily threefold: 

Firstly, we introduce an end-to-end unsupervised reconstruction framework tailored to the specific scanning conditions. It achieves HR 3D representation from motion-corrupted 2D stacks of fetal brain MR images by updating SVR and SRR networks in an iterative manner.
Secondly, we propose an unsupervised SVR method that employs a CNN to predict spatial transformations for aligning randomly positioned fetal brain MR slices into 3D target space. It is compatible with different resolutions and slice thicknesses, enabling robust alignment in real clinical data.
Thirdly, it includes an unsupervised SRR method that utilizes deep decoding network to learn an HR volume representation from input brain slices. Combining with the physical forward model, the proposed network incorporates a local consistency prior that reduces the problem of overfitting.
This work extends our previous research (ASSURED \cite{wu2023assured}) with several notable advancements. First, we have implemented an iterative training method alongside SVR-net, leading to more robust and accurate image reconstructions. Second, our introduction of a learnable intensity scale and outlier module significantly bolsters the reconstruction's robustness. Third, we have adopted a sampling point approach for PSF modeling, offering flexibility in handling various image sizes and enabling faster computational processes. These improvements collectively address the limitations identified in our earlier work, with a particular focus on enhancing reconstruction quality for fetuses with developmental abnormalities.


\begin{figure*}[!t]
\centerline{\includegraphics[width=\textwidth]{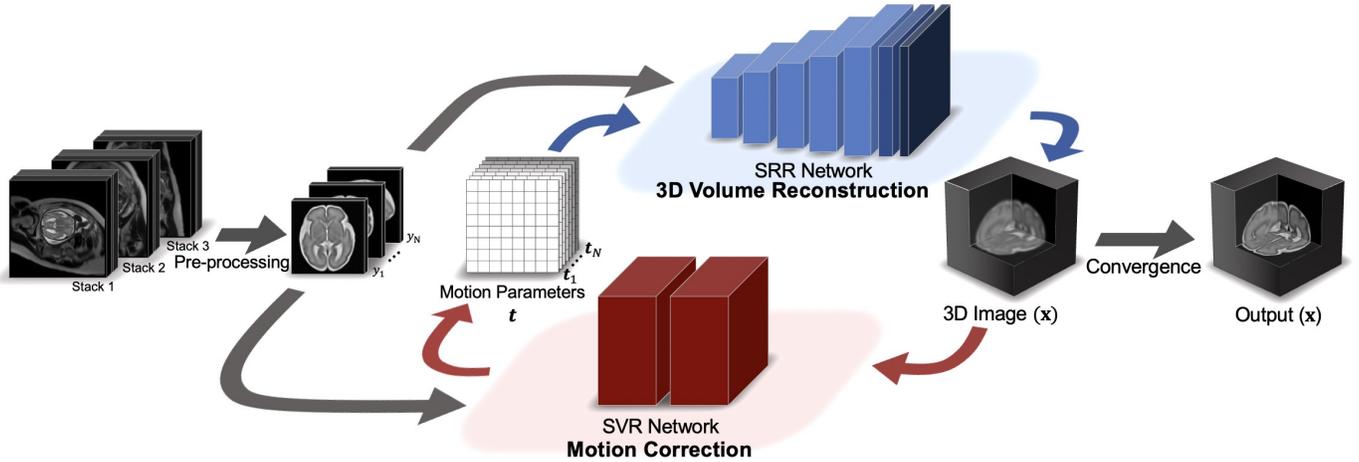}}
\caption{Our framework for fetal brain MRI reconstruction aims to achieve HR fetal brain volume from multiple motion-corrupted input LR stacks. The framework comprises an alternatingly optimized SVR-net and an SRR-net.
During training, the SVR-net initially aligns the motion-corrupted fetal brain stacks to a canonical 3D space and transfers the motion-corrected matrices to the SRR-net. The SRR-net, receiving the original stack of slices and these matrices, generates the 3D fetal brain volume from the motion-corrected MR slices. This reconstructed volume is then fed back to the SVR-net for further refinement of registration accuracy. Both networks undergo alternating optimization until convergence is achieved for the final high-resolution (HR) fetal brain output. In the inference stage, the high-resolution output from the SRR-net represents the final HR fetal brain image.
Details about SVR and SRR nets are presented in details in sections \ref{subsec:svrnet} and \ref{subsec:srrnet}, respectively.}
\label{pipeline}
\end{figure*}

\section{Related work}
\subsection{Motion correction for fetal brain MRI}
Fetal brain MRI motion correction usually utilizes rigid SVR to correct inter-slice motion. Numerous numerical optimization methods have been proposed for SVR, but their effectiveness depends on the degree of slice misalignment \cite{jiang2007mri,kim2009intersection,rousseau2006registration}. Conventional SVR struggles with significant slice displacement. In response, supervised DL-based methods have emerged for fetal brain SVR, predicting rigid transformations of MR slices through a data-driven mechanism \cite{hou2017predicting,salehi2018real,singh2020deep,shi2022affirm,xu2022svort}. Hou et al. \cite{hou2017predicting} employed a CNN to regress diverse rigid slice transformations for enhanced fetal brain MRI reconstruction. However, the registration accuracy diminishes when the SVR input image differs from the training data. Salehi et al. \cite{salehi2018real} proposed a conditional GAN-based image contrast transfer technique to enhance the generalization of the trained network. They trained a CNN model on T2-weighted fetal brain images, predicting 3D poses of newborn brains based on T1-weighted images. However, these methods process each slice independently, overlooking dependencies between slices. Singh et al. \cite{singh2020deep} introduced a recurrent network for inter-slice motion prediction in fetal MRI. Xu et al. \cite{xu2022svort} utilized a transformer-based network with an attention mechanism to predict rigid transformations for all slices, achieving a higher reliable transformation offset for brain motion than conventional registration-based approaches. However, supervised SVR methods still depend on large, high-quality training datasets to train effective networks.

Unsupervised image registration methods have attracted significant interest recently. Such as VoxelMorph \cite{balakrishnan2019voxelmorph},  is an unsupervised 3D volume-to-volume registration DL framework.  
Its learning mechanism enables network training using only the individual subject data pair without any supervision labels. 
However, for fetal brain SVR, the dimensional disparity between 2D and 3D images, coupled with quality degradation during 2D slice scanning, challenges the direct application of VoxelMorph. To address this, our study introduces SVR-net, featuring a dual-branch encoder for effective handling of slice and volume inputs. This design captures the unique characteristics of each input type. The SVR-net incorporates a volume slicing process, considering physical degradation effects like slice thickness and PSF blurring. Utilizing a comprehensive physical forward model aims to mitigate slicing degradation effects and enhance registration accuracy.

\subsection{Fetal brain reconstruction}

The HR fetal brain reconstruction, aiming for an isotropic HR volume from 2D input stacks, was pioneered by Rousseau et al. \cite{rousseau2006registration} with the first iterative SVR-SRR framework for fetal brain MRI. However, their use of a narrow Gaussian kernel as a PSF for HR volume interpolation during SVR may introduce blurring artifacts. To address this, Jiang et al. \cite{jiang2007mri} employed multilevel B-splines with a thickness of 1 mm to finely adjust the interpolation and minimize blurring in thinner slices. Kim et al. \cite{kim2009intersection} suggested a modified Gaussian-weighted reconstruction for a slice thickness of 3 mm but faced challenges with thicker slices ($>$3 mm), resulting in blurry reconstructed images.


Recent super-resolution methods \cite{greenspan2009super} combined with SVR aim to reduce blurring effects in reconstructed fetal brain MR images. Gholipour et al. \cite{gholipour2010robust} introduced a mathematical model for SRR using 2D fetal brain MR slices, implementing a constraint on the volume to mitigate noise amplification and registration errors. Rousseau et al. \cite{rousseau2013btk} employed total variation regularization to preserve anatomical structures during SRR. Murgasova et al. \cite{kuklisova2012reconstruction} demonstrated the effectiveness of outlier rejection using an automatic technique based on expectation maximization (EM)-based statistics. Ebner et al. \cite{ebner2020automated} proposed a slice-level outlier rejection method based on similarity scores but highlighted its reliance on the quality of the initial reconstructed volume, as significant corruption in the initial SRR volume can introduce errors in subsequent SVR and outlier removal steps.

One effective solution for reconstructing volumes from thick slice data involves improving the SRR process to provide a good initial volume for iterative SVR-SRR. McDonagh et al. \cite{mcdonagh2017context} proposed a context-sensitive technique using CNNs to upsample one of the multiple LR input stacks, serving as the initial configuration for SVR-SRR. Challenges remain in obtaining extensive auxiliary training datasets with HR image dependencies and addressing generalization issues.

The DIP framework, introduced for unsupervised learning in tasks like super-resolution, inpainting, and denoising \cite{ulyanov2018deep}, randomly initializes network inputs and weights without training datasets. During training, parameters are optimized by enforcing consistency with the corrupted image, projecting it onto the CNN-generatable image space. We develop a training dataset-free unsupervised SRR method using DIP-structure. Our approach features an under-parameterized deep decoder, enforcing image consistency priors during SRR to mitigate network overfitting from slice misalignment and motion artifacts. Integrating a physical forward model enhances robustness in reconstructing fetal brain MR volumes from thick slices ($\sim$6 mm) obtained during scanning.

%% file: LaTeX/section/method.tex
\section{Method}

\subsection{Overview of SUFFICIENT}
\label{subsec:Overview of SUFFICIENT}
The entire SUFFICIENT framework is depicted in Fig. \ref{pipeline}. It comprises a data preprocessing step (Details in section \ref{subsubsec:preprocessing}) and a fetal brain reconstruction step (Sections \ref{subsec:svrnet} \& \ref{subsec:srrnet}). The preprocessing step extracts motion-corrupted fetal brain stacks from three orthogonal orientations of uterine MRI and perform bias field correction, volume-to-volume registration, and intensity correction operations. Then the preprocessed image stacks are taken as input to the fetal brain reconstruction step, which iteratively performs motion correction (SVR-net) and 3D image super-resolution (SRR-net) to reconstruct an HR fetal brain volume. Specially, for all slices $\left\{\mathbf{y}_k\right\}_{k=1}^N,(k=1, \ldots, N)$ in the input stacks, we first predict their positions in a coronal fetal brain space and produce the rigid motion correction parameters $\boldsymbol{t}=$ $\left\{\boldsymbol{t}_k\right\}_{k=1}^N$ using the proposed SVR-net. Then we train an SRR model to reconstruct the 3D volume $\mathbf{x}$ from the motion-corrected slices. Both SVR and SRR networks are trained in an unsupervised manner. Sequentially, the iterative SVR-SRR process updates the motion parameters $\boldsymbol{t}$ and refines the fetal brain volume $\mathbf{x}$ until it converges.

\begin{figure}[!t]
\centerline{\includegraphics[width=\columnwidth]{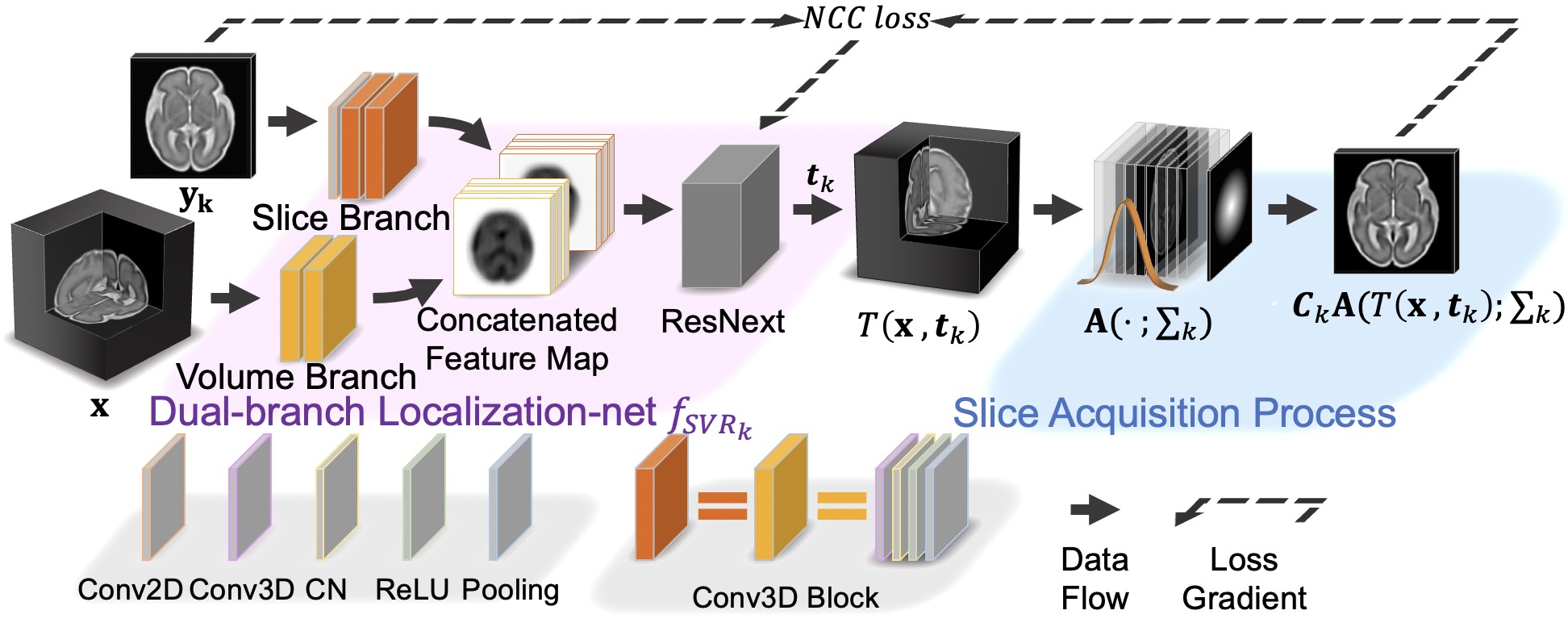}}
\caption{The proposed unsupervised SVR network for fetal brain motion correction. It predicts the transform matrix $\boldsymbol{t}_k$ that aligns the input slice $\mathbf{y}_k$ to 3D volume $\mathbf{x}$. The SVR-net comprises a dual-branch localization-net, and a differentiable slice acquisition model. During training, the network minimizes the error between the input slice $\mathbf{y}_k$ and the estimated slice $\hat{\mathbf{y}}_k$, which is computed using the slice acquisition model $\hat{\mathbf{y}}_k=$ $C_k\mathbf{A}\left(T\left(\mathbf{x}, \boldsymbol{t}_k\right) ; \Sigma_k\right)$ (E.q \ref{equation:forward}). }
\label{svr_pipeline}
\end{figure}

\subsection{Formulation of the Forward Model for MRI Slicing Process}
\label{subsec:forwardmodel}
To improve the performance of the inverse process in reconstructing HR 3D fetal brain MRI from LR 2D image stacks, we formulate the slicing process forward model as an MR slice acquisition model \cite{ebner2020automated,gholipour2010robust,kainz2015fast,kuklisova2012reconstruction}. Given a number of $N$ motion-corrupted slices $\left\{\mathbf{y}_k\right\}_{k=1}^N, k \in(1, \cdots, N)$, the forward model is defined as

\begin{equation}
\label{equation:forward}
\mathbf{y}_k=C_k \mathbf{A}\left(T\left(\mathbf{x}, \boldsymbol{t}_k\right) ; \Sigma_k\right)+\boldsymbol{\varepsilon}_k, k=1,2, \ldots, N
\end{equation}

where $\mathbf{x}$ is the $3 \mathrm{D}$ isotropic HR volume. $T(*)$ denotes a rigid transform that aligns the $2 \mathrm{D}$ image slice into its position in the $3 \mathrm{D}$ volume $\mathbf{x}$, guiding by the transform matrix $\boldsymbol{t}_k$. $C_k$ represents an intensity scaling operation applied to slice $k$, introduced to compensate for global intensity inhomogeneities across slices. This operation ensures that each slice $k$ is adjusted for consistent intensity levels relative to the entire volume.
$\mathbf{A}(*)$ denotes the MRI slicing operation, which is simulated with image downsampling and PSF blurring effect at the $k_{t h}$ slicing position \cite{liang2000principles}. $\boldsymbol{\varepsilon}_k$ represents additive noise that follows a Gaussian distribution \cite{gudbjartsson1995rician}. For SSFSE sequences, a common approximation of the PSF is given by a 3D Gaussian function defined by the variance-covariance matrix $\Sigma_k=$ $\operatorname{diag}\left(\frac{\left(1.2 s_1\right)^2}{8 \ln (2)}, \frac{\left(1.2 s_2\right)^2}{8 \ln (2)}, \frac{s_3^2}{8 \ln (2)}\right)$, where $s_1, s_2$ is the in-plane resolution and $s_3$ is the through-plane resolution \cite{kuklisova2012reconstruction}. The intensity of each voxel in an LR slice $\mathbf{y}_k$ is weighted by its neighbor voxels within the HR volume $\mathbf{x}$ using the defined PSF. Thus $\mathbf{A}(*)$ acts as an oriented Gaussian interpolation operation. For each voxel location $p$ in slice $\mathbf{y}_k$, the intensity of $p$ is computed as:
\begin{equation}
\mathbf{y}_k(p)=C_k\mathbf{A}\left(T\left(\mathbf{x}, \boldsymbol{t}_k\right) ; \Sigma_k\right)(p)+\boldsymbol{\varepsilon}_k(p)
\end{equation}
For the PSF blurring effect and image down-sampling operation, we need to consider the neighborhood voxels of $p$ in the volume $\mathbf{x}$. Specifically, we generate a number of $S$ independent samples from a Gaussian distribution, defined as $p^{\prime}=u_s+p$ with $u_s \sim \mathcal{N}(\mathbf{0}, \Sigma)$, for $s=1, \ldots, S$. The slicing process is then described by the following equations:
\begin{equation}
\mathbf{A}\left(\mathbf{m} ; \Sigma_k\right)(p)=\frac{1}{S} \sum_{s=1}^S g\left(u_s, \Sigma_k\right)\mathbf{m}(p^{\prime}_s)
\end{equation}
\begin{equation}
g\left(\mathbf{u}, \Sigma_k\right)=\frac{1}{\sqrt{\left.(2 \pi)^3 \operatorname{det}\left(\Sigma_k\right)\right)}} \exp \left(-\frac{1}{2} \mathbf{u}^T \Sigma_k^{-1} \mathbf{u}\right)
\end{equation}
where $\mathbf{m}=T\left(\mathbf{x}, \boldsymbol{t}_k\right)$ denotes the image slice located by $\boldsymbol{t}_k$. This forward model is differentiable, we can backpropagate errors during the network optimization process. In this work, the forward model is involved in both SVR and SRR nets.

\subsection{Development of the Unsupervised SVR-net}
\label{subsec:svrnet}
Fig. \ref{svr_pipeline} shows the SVR network training process. We train a function $\boldsymbol{t}_k=$ $f_{{SVR}_k}\left(\mathbf{y}_k, \mathbf{x}\right)$ to predict the transform matrix $\boldsymbol{t}_k$ that aligns each input 2D slice $\mathbf{y}_k$ towards the target 3D space $\mathbf{x}$, where $k=1, \ldots, N$ denotes the total number of slices from all input stacks. As well, $k$ is also an indicator that indices the slice selection position. Using the predicted transformation $\boldsymbol{t}_k$, we can transform the 3D volume $\mathbf{x}$ to the slicing position of the input slice $\mathbf{y}_k$ via $T\left(\mathbf{x}, \boldsymbol{t}_k\right)$. We then simulate the physical image degradation during slice acquisition process by a forward $\operatorname{model} \mathbf{A}(*)$, which includes image down-sampling and PSF blurring effect as presented in Section \ref{subsec:forwardmodel}. Then the $k_{t h}$ estimated slice $\hat{\mathbf{y}}_k=C_k\mathbf{A}\left(T\left(\mathbf{x}, \boldsymbol{t}_k\right) ; \Sigma_k\right)$ is computed from the current brain volume $\mathbf{x}$. Subsequently, we optimize the function $f_{S V R_k}$ by minimizing the difference between the predicted slice $\hat{\mathbf{y}}_k$ and the input slice $\mathbf{y}_k$.

\subsubsection{Dual-branch Localization-net}
The localization-net trains a function $f_{S V R_k}\left(\mathbf{y}_k, \mathbf{x}\right)$ that predicts the alignment parameter $\boldsymbol{t}_k$ between each input $\mathbf{y}_k$ and $\mathbf{x}$. It utilizes a CNN architecture, which includes a dual-branch feature extraction block and a ResNext network \cite{xie2017aggregated}. The 3D embedding branch is employed to extract image features from $\mathbf{x}$, while the 2D embedding branch is used for $\mathbf{y}_k$. The resulting feature maps from both branches are concatenated along the depth dimension and fed into the ResNext network to estimate the transformation parameter.
\subsubsection{Unsupervised SVR with a differentiable slice acquisition model}
Given a 2D slice $\mathbf{y}_k, k \in(1, \cdots, N)$, and a 3D volume $\mathbf{x}$ as inputs, we seek a mapping function $f_{S V R_k}$ that predicts the optimized motion parameter $\boldsymbol{t}_k^*$ by minimizing the objective function:
\begin{equation}
\boldsymbol{t}_k^*=\arg \max _{\boldsymbol{t}_k} \operatorname{Sim}\left(\mathbf{y}_k, \hat{\mathbf{y}}_k\right), \quad \boldsymbol{t}_k=f_{S V R_k}\left(\mathbf{y}_k, \mathbf{x}\right)
\end{equation}
where $\operatorname{Sim}(\cdot)$ is the matching indicator that quantifies the image similarity between the input slice $\mathbf{y}_k$ and the estimated slice $\hat{\mathbf{y}}_k$. In our study, we use the normalized cross-correlation (NCC) loss to estimate the similarity. Therefore, the proposed SVR network aligns slice $\mathbf{y}_k$ towards volume $\mathbf{x}$ by optimizing the loss function $\mathcal{L}_{SVR}$:
\begin{equation}
\mathcal{L}_{SVR}=NCC\left(\mathbf{y}_k,C_k\mathbf{A}\left(T\left(\mathbf{x} , \boldsymbol{t}_k\right) ; \Sigma_k\right)\right)
\end{equation}


\begin{figure}[!t]
\centerline{\includegraphics[width=\columnwidth]{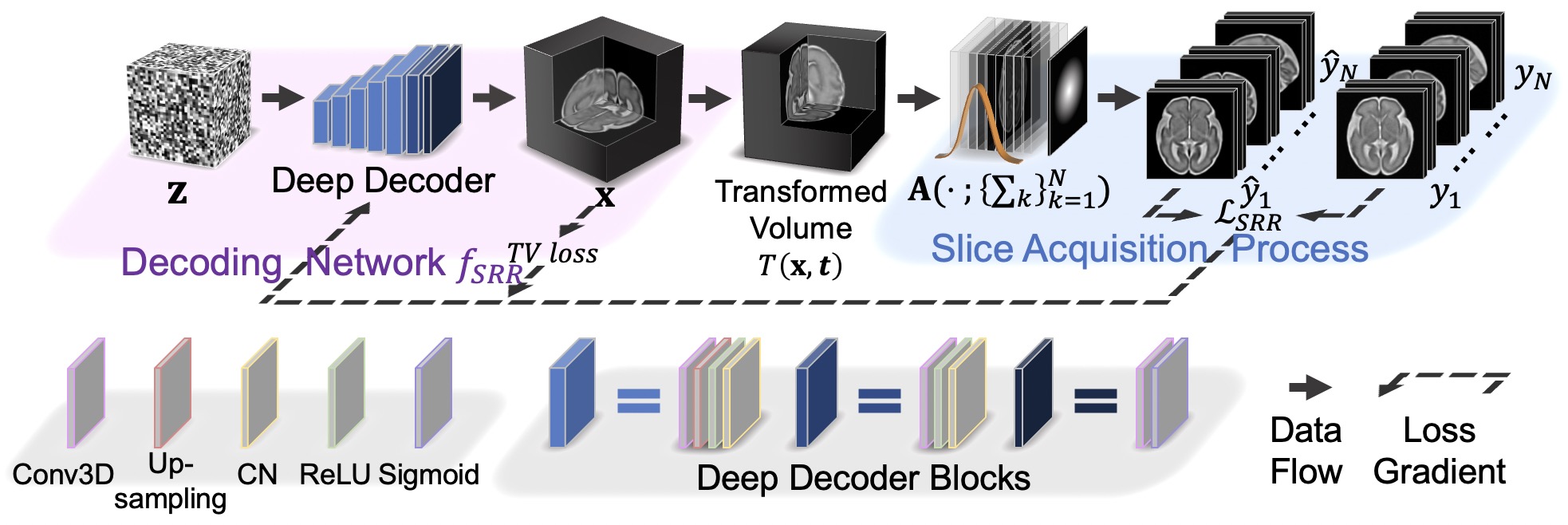}}
\caption{The proposed unsupervised SRR network for $3 \mathrm{D}$ volume reconstruction. It consists of a decoding network and a slice generation process. The decoding network takes a fixed noise $\mathbf{z}$ as input and maps it to an HR volume $\mathbf{x}$, which represents the fine fetal brain reconstruction from input motion-corrected brain slices. The motion correction for input slices is conducted using transformation matrix $t$ predicted by SVR-net. The SRR network is optimized by minimizing the difference between the predicted slices $\left\{\hat{\mathbf{y}}_k\right\}_{k=1}^N$ and the input slices $\left\{\mathbf{y}_k\right\}_{k=1}^N$, where $\hat{\mathbf{y}}_k=C_k\mathbf{A}\left(T\left(\mathbf{x}, \boldsymbol{t}_k\right) ; \Sigma_k\right)$ computed from the slice generation model in section \ref{subsec:forwardmodel}.}
\label{srr_pipeline}
\end{figure}

\subsection{Development of the Unsupervised SRR-net}
\label{subsec:srrnet}

Fig. \ref{srr_pipeline} shows the proposed SRR-net. We train a function $\mathbf{x}=f_{S R R}(\mathbf{z})$ to generate the $\mathrm{HR}$ reconstruction volume $\mathbf{x}$, where $\mathbf{z}$ is a fixed noise. Using the motion parameters $\boldsymbol{t}$ predicted by SVR-net, we can transform the 3D volume $\mathbf{x}$ to the slicing positions of all the input slices $\left\{\mathbf{y}_k\right\}_{k=1}^N$ by $T(\mathbf{x}, \boldsymbol{t})$. We then simulate the physical image degradation during slice acquisition using the forward MRI slice acquisition model $\mathbf{A}(*)$. Then the estimated slices $\hat{\mathbf{y}}_k=C_k\mathbf{A}\left(T\left(\mathbf{x}, \boldsymbol{t}_k\right) ; \Sigma_k\right), k=1, \cdots, N$ are generated from the volume $\mathbf{x}$. We optimize the function $f_{SRR}$ by minimizing the difference between the predicted slices $\left\{\hat{\mathbf{y}}_k\right\}_{k=1}^N$ and the input slices $\left\{\mathbf{y}_k\right\}_{k=1}^N$.

Specially, the HR volume $\mathbf{x}$ is represented by a randomly initialized deep decoder network $f_{S R R}$ defined by a set of parameters $\boldsymbol{\theta}$ with a fixed noise tensor $\mathbf{z}$. To improve the robustness of the learning process and to prevent the learning procedure from yielding local optima, random noise $v$ is incorporated into the network input in combination with $\mathbf{z}$ \cite{ulyanov2018deep,cheng2019bayesian}. The noise $v \sim \mathcal{N}\left(0, \sigma^2\right)$ is a Gaussian noise. The learning process for $f_{S R R}$ is then formulated by
\begin{equation}
\underset{\boldsymbol{\theta}}{\arg \min } \sum_{k=1}^N\left\|\mathbf{y}_k-C_k\mathbf{A}\left(T\left(f_{S R R}(\mathbf{z}+\boldsymbol{v}), \boldsymbol{t}_k\right) ; \Sigma_k\right)\right\|_2
\end{equation}
SRR is a challenging problem characterized by high under-determinacy and ill-posedness. To enhance the quality of the reconstructed images, regularization methods are employed. In this study, we utilize total variation as a regularization technique to improve local intensity consistency in the reconstructed volume. The reconstruction loss function $\mathcal{L}_{S R R}$ is thus further updated as follows:
\begin{equation}
\begin{split}
\mathcal{L}_{S R R}=&\sum_{k=1}^N\left\|\mathbf{y}_k-C_k\mathbf{A}\left(T\left(f_{S R R}(\mathbf{z}+\boldsymbol{v}), \boldsymbol{t}_k\right) ; \Sigma_k\right)\right\|_2\\
&+\lambda \ell_{T V}\left(f_{S R R}(\mathbf{z}+\boldsymbol{v})\right)
\end{split}
\label{equation:SRRloss}
\end{equation}
where $\|\cdot\|_2$ denotes the $L_2$ norm. $\ell_{T V}$ is the total variation, and $\lambda$ is a tuning parameter for the regularization term. 
To further enhance the robustness of the SRR against artifacts arising from motion corruption, we have incorporated a learnable slice outlier-rejection mechanism. This approach involves assigning a weight parameter $w_k$ to each slice $\mathbf{y}_k$. This parameter is designed to mitigate the influence of corrupted slices on the reconstruction outcome by dynamically adjusting their impact during the reconstruction process. The integration of $w_k$ effectively refines our reconstruction loss function, $\mathcal{L}_{S R R}$, which is now formulated as:
\begin{equation}
\begin{split}
\mathcal{L}_{S R R}=&\sum_{k=1}^N\left(\frac{\left(\mathbf{y}_k-\hat{\mathbf{y}}_k\right)^2}{2 w_k^2}
+\frac{1}{2} \log \left(w_k^2\right)\right)
\\&+\lambda \ell_{T V}\left(f_{S R R}(\mathbf{z}+\boldsymbol{v})\right)
\end{split}
\label{equation:SRRloss}
\end{equation}
    where $\hat{\mathbf{y}}_k = C_k\mathbf{A}\left(T\left(f_{S R R}(\mathbf{z}+\boldsymbol{v}), \boldsymbol{t}_k\right) ; \Sigma_k\right)$


%% file: LaTeX/section/exp.tex
\section{Experiments and results}
\subsection{Evaluations for estimating the efficiency of the proposed model}
To comprehensively evaluate the proposed SUFFICIENT model, we conduct the following four experiments:
\begin{enumerate}[label=(\roman*)]
    \item \textbf{\textit{Fetal brain reconstruction on the simulated dataset with different levels of motion corruption:}} The performance of fetal brain reconstruction heavily depends on the level of fetal motion during MR scanning. In this experiment, we evaluated the SUFFICIENT model and four baseline models on a simulated dataset with varying levels of motion corruption (e.g., low-level and high-level). 
    \item \textbf{\textit{Fetal brain reconstruction task on the real clinical dataset:}} Clinical MR datasets often exhibit different contrast levels and more realistic artifacts, noise, and fetal motion compared to synthetic data. In this experiment, we evaluated the SUFFICIENT model and four baseline models on a clinical dataset. 
    \item \textbf{\textit{Efficiency of SVR-Net:}} The accuracy and generalization ability of SVR for different resolutions, thicknesses, and motion levels are critical factors affecting fetal brain reconstruction performance. In this experiment, we compared it with three SVR methods on two simulation datasets with varying input characteristics. Specifically, we employed SVR-Net and three SVR components of baseline methods to perform motion correction on the simulation dataset, where all slices were registered to the GT. We calculated the difference between the predicted and true motion parameters.
    \item \textbf{\textit{Efficiency of SRR-Net:}} The performance of SRR is susceptible to motion artifacts and image noise. In this experiment, we compared it with three SRR methods on a simulation dataset where motion was coarsely corrected by the SVR method. We employed SRR-Net and three SRR components of baseline methods to perform brain volume reconstruction with only an initial SVR step. 
\end{enumerate}

\subsection{Experimental Settings}
\subsubsection{Simulated dataset construction}\hfill

\textbf{Ground truths data:} Ground truths (GTs) for 3D isotropic HR fetal brain MR images are typically unavailable. To evaluate the proposed network, we utilized fetal brain atlases  \cite{wu2021age} as the GT images and generated simulated 2D image stacks with motion corruption. Fourteen fetal brain atlas images with a resolution of 0.8 mm isotropic were considered as GTs.

\textbf{Different levels of motion corruption:} We denote the fetal brain motion correction parameter by a 3D rigid transform matrix $t_k$ with 6 degrees of freedom (DoF), including three rotation parameters $\boldsymbol{\alpha}=\left(\alpha_x, \alpha_y, \alpha_z\right)^T$ and three translation parameters $\boldsymbol{d}=\left(d_x, d_y, d_z\right)^T$, in the 3D space. The rotation component, $\boldsymbol{R} \in \mathbb{R}^{3 \times 3}$, is derived from the rotation parameters $\boldsymbol{\alpha}$. The complete rigid transformation, encompassing both rotation and translation, is represented in a $4 \times 4$ matrix format for practical implementation:
    $$
    t_k(\boldsymbol{R}, \boldsymbol{d})=\left[\begin{array}{cc}
    \boldsymbol{R} & \boldsymbol{d} \\
    0 & 1
    \end{array}\right]
    $$     
    To simulate motion-corrupted slices, we adopted a control-point-based scheme, which generates a continuous motion trajectory based on several control points using a simple random walk model \cite{fogtmann2013unified,singh2020deep,shi2022affirm}. In this study, the random offset parameters (3 for $\boldsymbol{\alpha}$ and 3 for $\boldsymbol{d}$ ) are sampled from uniform distribution $U(-i, i)$, where $[-i, i]$ denotes a variation range following the uniform distribution. Higher values of $i$ indicate the slice was affected by more severe brain motion. The motion parameters $\boldsymbol{t}$ were then applied to the GT brain atlas volume to generate simulated motion-corrupted MR image stacks, using the physical slice acquisition model described in section \ref{subsec:forwardmodel}.  

For all simulation data, we added random Gaussian noise to each slice, with zero mean and a standard deviation of $5 \%$ of the maximum voxel intensity of the image \cite{gholipour2010robust,sui2022scan}. In-plane motion artifacts were simulated following the approach described in \cite{perez2021torchio}. The specific details of the datasets are summarized in Table \ref{table_dataset}. Each dataset includes 42 motion-corrupted stacks from 14 GT brain atlases (each GT data generated 3 orthogonally oriented tacks). 

\begin{figure}[!]
\centerline{\includegraphics[width=\columnwidth]{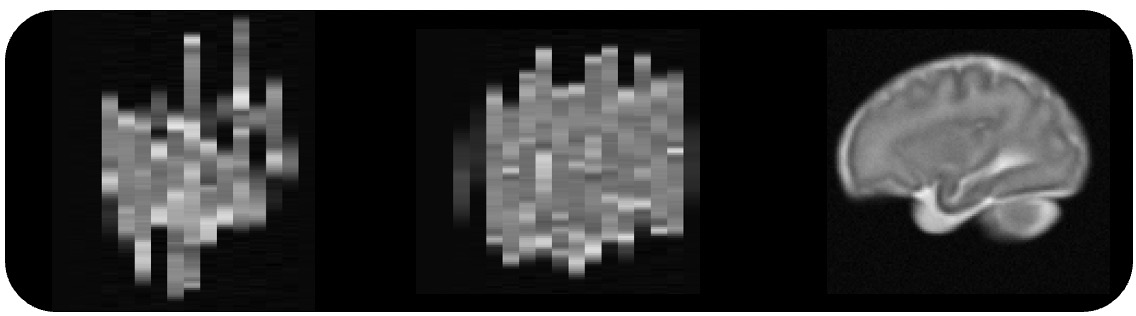}}
\caption{A representative simulated stack sample from Group B in coronal (left), axial (mid) and sagittal (right) views.}
\label{simulation_sample}
\end{figure}

\input{LaTeX/table/table_dataset}

\subsubsection{Real clinical fetal brain dataset}The real clinical dataset used in this study was retrospectively collected from archived MRI data. The data were acquired on a $1.5 \mathrm{~T}$ scanner (Siemens Magnetom Aera Syngo MR E11) with eight-channel body coil, using weighted half-Fourieracquisition (T2w) SSFSE sequence. The scanning parameters were set as: TR/TE = 1400/92 ms, in-plane resolution =1.172 $\times$ 1.172 mm, and slice thickness of 6-6.5 mm. The clinical experimental dataset consisted of T2-weighted MR images from 24 fetuses with gestational ages ranging from 22 weeks to 35 weeks. Furthermore, to augment the robustness of our validation, the study was extended to incorporate additional clinical datasets encompassing 12 fetuses with slice thicknesses ranging from 3–4 mm.

\subsubsection{Data preprocessing for clinical fetus MRI data}
\label{subsubsec:preprocessing}
The data preprocessing conforms to the following procedure. First, we utilized a trained segmentation network  \cite{ebner2020automated} to extract the fetal brains from the uterine MRI. Next, we applied bias field correction \cite{tustison2010n4itk} to eliminate the bias field present in each slice. Volume-to-volume registration \cite{kainz2015fast} using NiftyReg \cite{modat2014global} was employed to align and consolidate the spatial information of stacks that were scanned in different orientations. Following the volumetric alignment of stacks, we implemented an intensity normalization procedure to correct the intensities of each stack based on the target stack. The process involves calculating an initial intensity scaling factor, $C_i$, for each slice $\mathbf{y}_i$ in the comparative stack. This scaling factor is determined through a linear regression optimization, aimed at minimizing the intensity differences between the target stack and the current stack. The optimization equation used for this purpose is presented as follows:
\begin{equation}
\underset{\boldsymbol{C_i}}{\arg \min } \left\| \text { stack }_{\text {target }}-\sum_{i=1}^{N_s} C_i \mathbf{y}_i \right\|_2
\end{equation}
Here, $\text { stack }_{\text {target }}$ represents the intensity profile of the target stack, and $N_s$ is the total number of slices in one stack being normalized. By applying this linear regression approach, we ensure that the intensity scale of all stacks is initialized to a standard reference.

\begin{figure*}[!t]
\centerline{\includegraphics[width=\textwidth]{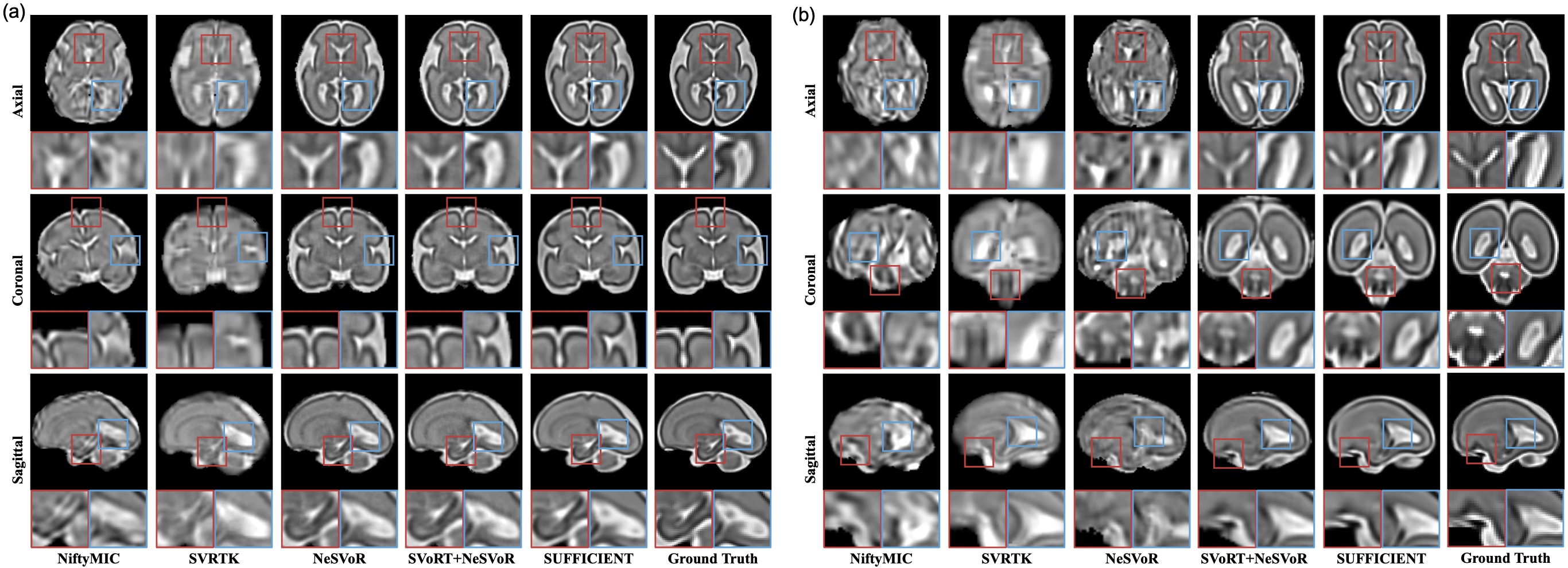}}
\caption{(a) The qualitative reconstruction results obtained on the Group A dataset of a representative subject at a 27-week GA. (b) The qualitative reconstruction results obtained on the Group B dataset of a representative subject at a 23-week GA.}
\label{groupBC}
\end{figure*}

\subsubsection{Methods for comparison} We selected four SOTA fetal brain reconstruction methods that provided open-source implementations for comparison purposes. These methods are as follows: 1) SVRTK: a classic slice-to-volume fetal brain reconstruction toolkit \cite{kuklisova2012reconstruction}, 2) NiftyMIC: an automatic fetal brain reconstruction framework \cite{ebner2020automated}, 3) NeSVoR: a slice-to-volume reconstruction method based on implicit neural representation \cite{xu2023nesvor}, and 4) SVoRT+NeSVoR: an approach utilizing a pretrained SVR method, SVoRT \cite{xu2022svort}, to provide initialized transformations for NeSVoR. To ensure fair comparisons, we conducted hyperparameter tuning for the comparison methods. Specifically, we randomly picked one subject from the simulated fetal dataset and tuned the hyperparameters to minimize the MSE between the reconstructed volume and the GT. Once the hyperparameters were determined, they were fixed and applied to both the simulated dataset and the real clinical dataset in the comparison experiments.

\subsubsection{Evaluation metrics}The simulated fetal data was reconstructed with an isotropic resolution of 0.8 mm to match the original resolutions of the ground truth images. We conducted a comparison between the reconstructed volumes and GT using various quantitative metrics, including the peak signal-to-noise ratio (PSNR), structural similarity index measure (SSIM) \cite{zhao2019channel}, and NCC.

\subsubsection{Training details}
The training process was conducted on a single NVIDIA TITAN RTX GPU. In the iterative SVR-SRR process, we performed one SVR step every 500 SRR epochs, resulting in a total of 8000 epochs. 

\subsection{Results and comparison on the simulation dataset}
\label{subsec:simulation_result}

\begin{figure*}[!t]
    \centerline{\includegraphics[width=\textwidth]{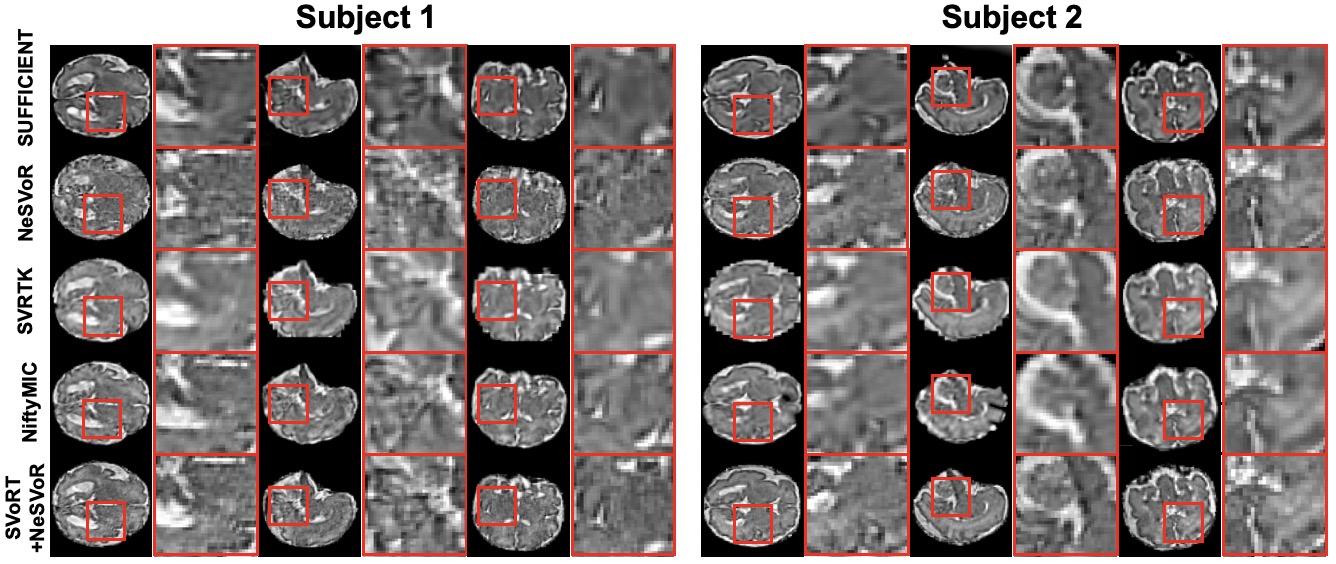}}
    \caption{Qualitative comparison of different reconstruction methods on two representative fetal brain subjects. The figure showcases results from individual spaces. The red boxes highlight areas of interest for closer examination. }
    \label{clinical_6mm_subject}
    \end{figure*}

\input{LaTeX/table/table_ABC}

Table \ref{table_ABC} presents the quantitative results of the comparison. Across all evaluation metrics, the SUFFICIENT method achieves the highest performance. The qualitative results of fetal brain reconstruction using different methods on the Group A dataset are presented in Fig. \ref{groupBC}a. The motion level in Group A is moderate, and only SUFFICIENT, NeSVoR, and SVoRT+NeSVoR achieve satisfactory reconstruction performance. Specifically, both SVRTK and NiftyMIC results are significantly affected by motion artifacts and image noise, resulting in inadequate volume integrity. In the coronal view, SUFFICIENT, NeSVoR, and SVoRT+NeSVoR exhibit clear cortical structure details. In the axial and sagittal views, SUFFICIENT outperforms the other two methods in terms of clearer ventricle boundaries and less global volume noise. Fig. \ref{groupBC}.b presents the qualitative results of fetal brain reconstruction comparison on the Group B dataset. The motion level in Group B is severe, and the results obtained from SVRTK, NiftyMIC, and NeSVoR are significantly affected by motion artifacts, resulting in insufficient volume integrity. The results from SUFFICIENT and SVoRT+NeSVoR exhibit better recovery of cortical structure details in the coronal view. In the axial and sagittal views, the structure of the ventricle in the results produced by SUFFICIENT are clearer and more consistent with the GT (as observed in the close-up view in the blue box). Conversely, the results of SVoRT+NeSVoR display inconsistencies with the GT in brain volume boundaries and local structure (as seen in the close-up view in the red box), and the global volume appears noisier compared to SUFFICIENT.



    \begin{figure*}[htbp]
    \centerline{\includegraphics[width=\textwidth]{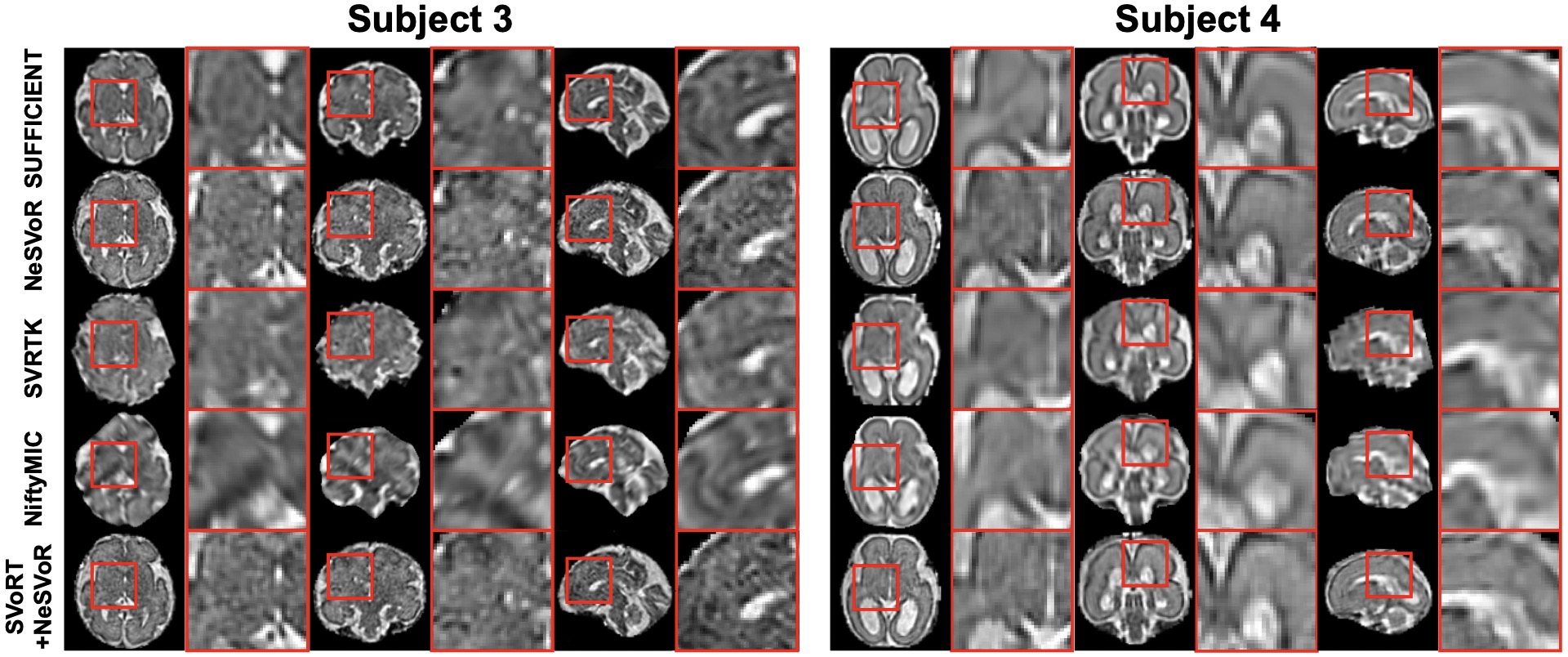}}
    \caption{Qualitative comparison of different reconstruction methods on two representative fetal brain subjects, illustrating reconstruction results in the standard radiological anatomical plane. The red boxes emphasize the detailed structures and contrasts between brain compartments.}
    \label{clinical_6mm_template}
    \end{figure*}

\subsection{Reconstruction results on clinical data}
\label{subsec:clinical_result}

Fig. \ref{clinical_6mm_subject} illustrates the qualitative results of fetal brain reconstruction on two motion-corrupted subjects in the individual space (original fetal brain position during MR scanning). The reconstructed volumes using SUFFICIENT demonstrate superior results regarding volume integrity, noise suppression, and local regional intensity consistency. These subjects experienced severe motion corruption and noise during scanning, posing challenges for accurate brain reconstruction. The results from SVRTK, NiftyMIC, and NeSVoR are significantly affected by motion artifacts, resulting in compromised volume integrity. SVoRT+NeSVoR's reconstructions, while somewhat effective, suffered from notable local regional intensity inconsistencies and noise, particularly evident in the detailed views within the red boxes. These areas appear noisy, making it difficult to discern the contrast between key brain structures. In contrast, SUFFICIENT’s reconstructions allow for the clear visualization of anatomical details, such as the Putamen in the coronal view of subject 1 and the gray matter in the axial view of subject 2.\input{LaTeX/table/table_SVR}To further evaluate the reconstructed results in standard radiological anatomical planes, we rigidly aligned the volumes to a spatiotemporal atlas of normal brains \cite{wu2021age} using NiftyReg \cite{modat2014global}. Fig. \ref{clinical_6mm_template} presents the reconstructed results of two challenging motion-corrupted fetal subjects in the standard radiological anatomical plane. Consistent with previous observations, reconstructions by SVoRT+NeSVoR were marred by noticeable noise and local regional intensity inconsistency. In contrast, volumes reconstructed with the SUFFICIENT method showcased superior outcomes. These reconstructions maintained the integrity of fetal brain cortical structures, exhibited effective noise suppression, and upheld local regional intensity consistency across subjects. Notably, reconstructions using SUFFICIENT provided clearer delineation of critical anatomical structures. For example, the Putamen was more visible in the axial view of subject 3. Moreover, the distinction between white and gray matter was more apparent, offering enhanced clarity in both the axial and coronal views of subject 4.

\input{LaTeX/table/table_SRR}

\subsection{Ablation experiments}
\label{subsec:ablation_result}

\textbf{Ablation studies for SVR-net:} In order to assess the contributions of individual components in our proposed SUFFICIENT framework, we conducted ablation studies. Firstly, we evaluated the accuracy of our SVR approach by performing slice motion corrections using SVR-net and three SVR components from baseline methods (SVRTK, NiftyMIC, SVoRT) on the Group B dataset. We did not consider NeSVoR in this experiment as it does not have an independent SVR component. For this study, we registered all slices to the ground truth using different SVR methods and calculated the differences between the predicted motion parameters and the true motion parameters. Additionally, we performed experiments on the Group C dataset to demonstrate the robustness of SVR to varied slice thicknesses and in-plane resolution.

\textbf{Ablation studies for SRR-net:} Secondly, we evaluated the robustness of our SRR approach by using only the SRR-net and the SRR part of three baseline methods to perform brain volume reconstruction with a fixed SVR step on the Group C dataset, which unified the motion effects on all SRR methods in comparison for fairness. For NeSVoR, the transformation parameters were not allowed to be updated during the training. In this experiment, we first registered all slices to the fetal brain atlas using our SVR-net. Then, we applied our SRR-net and the three baseline SRR methods to reconstruct the fetal brain volume. Finally, we evaluated the similarity between the reconstruction results and the GTs.


Table \ref{SVR_table} presents the quantitative results of the comparison between our SVR approach and the three baseline methods in terms of motion estimation accuracy. The mean absolute error (MAE) and RMSE of motion parameters were used as evaluation metrics. Our method achieved the best performance, as evidenced by the lowest MAE values of 0.22 mm for translation estimation and 0.23 degrees for rotation estimation on the Group B dataset. On the Group C dataset, our approach also outperformed the baseline methods, with the lowest average MAE values of 0.32 mm for translation estimation and 0.99 degrees for rotation estimation. Table \ref{SRR_table} shows that our SRR method achieved the highest average PSNR/SSIM/NCC of 22.34/0.8225/0.9202, and the lowest average RMSE of 0.0825. These results confirm the effectiveness of our SRR method in producing high-quality reconstructed fetal brain volumes. 

\section{Discussion}
Reconstructing HR 3D volumes from motion-corrupted stacks of 2D thick slices is a very challenging task. In this study, we proposed a novel scan-specific unsupervised reconstruction framework called SUFFICIENT, which does not require auxiliary training datasets. SUFFICIENT utilizes unsupervised SVR and SRR networks to perform 3D volume reconstruction iteratively and reliably. Our experiments on large-magnitude motion-corrupted simulation data demonstrated that SUFFICIENT outperformed other state-of-the-art methods. Moreover, our approach exhibited improved robustness to motion corruptions, and also reduced motion artifacts and imaging noise on clinical data.
For motion correction, learning-based methods have become increasingly popular in fetal SVR research \cite{hou20183,salehi2018real,shi2022affirm,singh2020deep,xu2022svort}. However, due to the lack of 3D fetal brain scannings, most models typically rely on synthetic data for training. Limited by the common generalization issue of supervised learning paradigm, the registration performance is heavily influenced by the distribution of the training data. This challenge is further escalated by the rapid neurodevelopment during the second trimester. The Fast growth rate results in significant differences in fetal brain structures at adjacent gestational ages, which brings additional difficulty for supervised learning-based methods to learn all related variations from very limited datasets \cite{singh2020deep}. To address this, robust scan-specific methods that requiring no auxiliary training dataset are recently one of the most reliable algorithm development trends. Two of the most recent SOTA comparison methods NeSVoR and SVoRT+NeSVoR are both scan-specific unsupervised learning methods. In NeSVoR, the authors proposed to perform SRR via the recent popular implicit neural representation (INR) network, and conducted SVR by joint optimizing the trainable transformation matrix when the network training. Additionally, a physical MRI degradation model was proposed NeSVoR to assist the INR learning high-quality fetal brain from degraded MR thick slices. NeSVoR achieved decent and robust reconstruction performance and with an enhanced SVR module as in SVoRT+NeSVoR, the performance is further improved. However, compared with the proposed SUFFICIENT, we believe the drawback in NeSVoR is the powerful ability of INR network on image detail learning, which induced sensitivity and overfitting on motion artifact in the motion degraded slices. Inspired by the NeSVoR, we propose a scan-specific unsupervised SVR method that combines a physical forward model and a generative network with more learning bias towards low-frequency image content. We trained the deep decoder network (DIP) to progressively learn image global content to local details. The ablation study results on SRR-net proved the efficiency of our strategy. On the other hand, for avoiding overfit to motion artifact and imaging noise, SUFFICIENT sacrificed its ability to fine image texture details. Specifically, the high-frequency detail information, such as structure details in the cerebellum, are lost in the reconstruction.

The deep decoder structure, in combination with the physical forward model, can provide an MR image prior that improves local regional consistency and brain cortical consistency in the reconstructed brain volume. This enhancement could aid in cortical development studies and whole-brain quantitative analyses. Despite these achievements, the present work has several limitations. The reconstruction time of SUFFICIENT is relatively long. Future research will focus on reducing the reconstruction time while considering the sequential information present in the acquired slices.

\section{Conclusion}
In conclusion, our proposed SUFFICIENT framework has demonstrated promising results in high-resolution 3D isotropic fetal brain MRI reconstruction. By leveraging scan-specific unsupervised SVR and SRR networks, SUFFICIENT achieves effective 3D volume reconstruction without the need for auxiliary training datasets or supervision labels. The framework's performance was extensively validated through experiments on both motion-corrupted simulation data and clinical data, showcasing its superiority over four state-of-the-art methods. Furthermore, SUFFICIENT proves to be a practical and feasible solution for clinical applications, as it can handle fetal brain reconstruction across various image resolutions and slice thicknesses without being constrained by the settings in the training data. 




%% file: LaTeX/table/table_dataset.tex
\begin{table}
\centering
\caption{ The simulated datasets were used for the experiments. The motion parameters (rotation and translation parameters) and the number of subjects and stacks are listed.}
\resizebox{\columnwidth}{!}{
\begin{tabular}{ccc}
\hline                                                                      & Group A                   & Group B                                                       \\ \hline
$\begin{array}{c}\text { Translation offset range } \\\text { described in } \boldsymbol{d} (\mathrm{mm})\end{array}$                                            & $(-4,4)$                             & $(-9,9)$                                     \\
$\begin{array}{c}\text { Rotation offset range described } \\\text { in } \boldsymbol{\alpha} \text { (degree) }\end{array}$       & $\left(-4^{\circ}, 4^{\circ}\right)$ & $\left(-9^{\circ}, 9^{\circ}\right)$   \\
Inplane resolution $\left(\mathrm{mm}^2\right)$                             & 0.8                                  & 0.8                                                                    \\
Slice thickness (mm)                                                       & 6                                    & 6                                                                               \\
Number of subjects                                                          & 14                                   & 14                                                                              \\
$\begin{array}{c}\text { Total number of image stacks } \end{array}$ & 42                                   & 42                                                                        \\ \hline
\end{tabular}}

\label{table_dataset}
\end{table}

%% file: LaTeX/table/table_ABC.tex
\begin{table}[h]
\centering
\caption{The statistical results of quantitative metrics produced by different methods on the Group A and B datasets.}
\resizebox{\columnwidth}{!}{
\begin{tabular}{l|ccc}
\hline
\textbf{Method} & \textbf{PSNR} $\uparrow$ & \textbf{SSIM} $\uparrow$ & \textbf{NCC} $\uparrow$ \\ \hline

\multicolumn{4}{c}{\textbf{Group A Dataset}} \\ \hline
SVRTK        & $16.86 \pm 1.75$   & $0.6450 \pm 0.1160$   & $0.7430 \pm 0.0955$ \\
NiftyMIC     & $15.63 \pm 4.14$   & $0.6757 \pm 0.2327$   & $0.6300 \pm 0.2219$ \\
NeSVoR       & $21.21 \pm 4.03$   & $0.8174 \pm 0.1063$   & $0.8848 \pm 0.0833$ \\
SVoRT+NeSVoR & $20.99 \pm 4.62$         & $0.8092 \pm 0.0853$                           & $0.8725 \pm 0.0825$                 \\
\textbf{SUFFICIENT} & $\mathbf{22.39 \pm 2.87}$ & $\mathbf{0.8810 \pm 0.0476}$ & $\mathbf{0.9278 \pm 0.0266}$ \\ \hline

\multicolumn{4}{c}{\textbf{Group B Dataset}} \\ \hline
SVRTK        & $16.12 \pm 1.68$   & $0.6029 \pm 0.0999$   & $0.6841 \pm 0.1247$ \\
NiftyMIC     & $14.74 \pm 3.59$   & $0.4617 \pm 0.2735$   & $0.4477 \pm 0.2881$ \\
NeSVoR       & $16.18 \pm 3.02$   & $0.6198 \pm 0.2416$   & $0.7049 \pm 0.2396$ \\
SVoRT+NeSVoR & $20.25 \pm 5.46$         & $0.7980 \pm 0.1881$                           & $0.8515 \pm 0.1970$                 \\
\textbf{SUFFICIENT} & $\mathbf{21.40 \pm 3.28}$ & $\mathbf{0.8718 \pm 0.0594}$ & $\mathbf{0.9033 \pm 0.0397}$ \\ \hline
\end{tabular}
}
\label{table_ABC}
\end{table}

%% file: LaTeX/table/table_SVR.tex
\begin{table*}
\centering

\caption{ Quantitative comparison of motion estimation accuracy between our approach and three SVR components (from SVRTK, NiftyMIC and SVoRT) of fetal brain reconstruction frameworks on the simulation dataset Group B and C. }
\resizebox{\textwidth}{!}{
\begin{tabular}{ccccccccc}
\hline
\multicolumn{1}{c|}{\multirow{3}{*}{Method}} & \multicolumn{4}{c}{Group B dataset}                                                                                                        & \multicolumn{4}{c}{Group C dataset}                                                                                         \\ \cline{2-9} 
\multicolumn{1}{c|}{}                        & \multicolumn{2}{c}{Translation $(mm)$}                       & \multicolumn{2}{c}{Rotation (degree)}                                       & \multicolumn{2}{c}{Translation $(mm)$}                       & \multicolumn{2}{c}{Rotation (degree)}                        \\ \cline{2-9} 
\multicolumn{1}{c|}{}                        & $\mathrm{MAE} \downarrow$ & $\operatorname{RMSE} \downarrow$ & $\mathrm{MAE} \downarrow$ & \multicolumn{1}{c|}{$\mathrm{RMSE} \downarrow$} & $\mathrm{MAE} \downarrow$ & $\operatorname{RMSE} \downarrow$ & $\mathrm{MAE} \downarrow$ & $\operatorname{RMSE} \downarrow$ \\ \hline
\multicolumn{1}{c|}{SVRTK-SVR}               & $0.79 \pm 0.51$           & $1.05 \pm 0.43$                  & $0.85 \pm 0.26$           & \multicolumn{1}{c|}{$1.09 \pm 0.50$}            & $4.64 \pm 1.17$           & $5.41 \pm 0.99$                  & $17.25 \pm 2.94$          & $17.32 \pm 4.28$                 \\
\multicolumn{1}{c|}{NiftyMIC-SVR}            & $0.63 \pm 0.24$           & $0.86 \pm 0.08$                  & $0.66 \pm 0.30$           & \multicolumn{1}{c|}{$0.88 \pm 0.29$}            & $4.38 \pm 1.17$           & $5.27 \pm 2.55$                  & $13.48 \pm 1.19$          & $16.32 \pm 1.31$                 \\
\multicolumn{1}{c|}{SVoRT}                   & $0.58 \pm 0.29$           & $0.87 \pm 0.59$                  & $0.62 \pm 0.23$           & \multicolumn{1}{c|}{$0.84 \pm 0.12$}            & $2.58 \pm 1.02$           & $3.19 \pm 0.11$                  & $6.52 \pm 1.42$           & $8.64 \pm 1.45$                  \\
\multicolumn{1}{c|}{Our SVR}                 & $\bold{0.22 \pm 0.16}$           & $\bold{0.29 \pm 0.17}$                  & $\bold{0.23 \pm 0.06}$           & \multicolumn{1}{c|}{$\bold{0.30 \pm 0.05}$}            & $\bold{0.32 \pm 0.30}$           & $\bold{0.64 \pm 0.58}$                  & $\bold{0.99 \pm 0.87}$           & $\bold{1.28 \pm 1.19}$                  \\ \hline                         
\end{tabular}}
\label{SVR_table}
\end{table*}

%% file: LaTeX/table/table_SRR.tex
\begin{table}
\centering
\caption{  Quantitative comparison results between our SRR approach and three baseline methods (from SVRTK, NiftyMIC and NeSVoR) on the simulation Group C dataset.}
\resizebox{\columnwidth}{!}{
\begin{tabular}{c|ccc}
\hline
Methods           & $\operatorname{PSNR} \uparrow$ & $\operatorname{SSIM} \uparrow$ & $\mathrm{NCC} \uparrow$       \\ \hline
SVRTK             & $18.33 \pm 3.77$               & $0.6779 \pm 0.1944$            & $0.7770 \pm 0.1550$      \\
NiftyMIC          & $19.06 \pm 3.56$               & $0.7393 \pm 0.1386$            & $0.8440 \pm 0.0882$      \\
$\mathrm{NeSVoR}$ & $19.48 \pm 2.87$               & $0.7656 \pm 0.1196$            & $0.8785 \pm 0.0736$      \\
Our SRR           & $\bold{22.34 \pm 3.70}$          & $\bold{0.8225 \pm 0.0935}$            & $\bold{0.9202 \pm 0.0457}$      \\ \hline
\end{tabular}}
\label{SRR_table}
\end{table}

%% file: LaTeX/section/acknowledge.tex
\section*{Acknowledgment}
This work was supported by the National Natural Science Foundation
of China under Grant 62071299, the Guizhou senior innovative talent project (grant numbers QKHPTRC-GCC[2022]041–1), and the Guizhou Provincial Science and Technology project (QKHZC[2019]2810).